%% file: main.arxiv.tex
\documentclass[a4paper,11pt]{article}
\usepackage{preamble}
\usepackage{natbib}
\usepackage{fullpage}

\title{\bf Estimating the Expected Social Welfare and Cost\\of Random Serial Dictatorship}

\author{
    Ioannis Caragiannis \and Sebastian Homrighausen
}
\date{Department of Computer Science, Aarhus University\\\{iannis, homrighausen\}@cs.au.dk
}

\begin{document}

\maketitle

\begin{abstract}
    \subfile{abstract}\label{sec:abstract}
\end{abstract}

\section{Introduction}\label{sec:introduction}
\subfile{introduction}

\section{Preliminaries}\label{sec:preliminaries}
\subfile{preliminaries}

\section{\#P-Hardness of Random Serial Dictatorship}\label{sec:hardness}
\subfile{hardness}

\section{Approximating the Expected Social Welfare}\label{sec:approxswelfareofrsd}
\subfile{approxsocialwelfare}

\section{Approximating the Expected Social Cost}\label{sec:approxscostofrsd}
\subfile{approxsocialcost}

\section{Conclusion}\label{sec:discussion}
\subfile{discussion}

\subsection*{Acknowledgments}
This work was supported by the Independent Research Fund Denmark (DFF) under grant 2032-00185B.

\bibliography{bibliography}

\appendix
\subfile{appendix}

\end{document}

%% file: abstract.tex
We consider the assignment problem, where $n$ agents have to be matched to $n$ items. Each agent has a preference order over the items. In the serial dictatorship (SD) mechanism the agents act in a particular order and pick their most preferred available item when it is their turn to act. Applying SD using a uniformly random permutation as agent ordering results in the well-known random serial dictatorship (RSD) mechanism. Accurate estimates of the (expected) efficiency of its outcome can be used to assess whether RSD is attractive compared to other mechanisms. In this paper, we explore whether such estimates are possible by sampling a (hopefully) small number of agent orderings and applying SD using them. We consider a value setting in which agents have values for the items as well as a metric cost setting where agents and items are assumed to be points in a metric space, and the cost of an agent for an item is equal to the distance of the corresponding points. We show that a (relatively) small number of samples is enough to approximate the expected social welfare of RSD in the value setting and its expected social cost in the metric cost setting despite the \#P-hardness of the corresponding exact computation problems. 

%% file: introduction.tex
We consider assignment problems in which a set of $n$ agents must be assigned (matched) to a set of $n$ items. In an assignment instance each agent has a preference ranking over the items. According to the most straightforward mechanism known as {\em serial dictatorship} (SD), the agents are asked to act in a predefined order, and when it is their turn to act, they are assigned to their favourite item that has not been selected by other agents in previous steps. 

SD achieves several desirable properties. For example, the interaction with the mechanism is minimal and intuitive from an agent's perspective. Also, SD is probably the most natural representative in the field of mechanism design without money~\citep{PT09,SV07}, and produces Pareto-efficient assignments when the agent preferences are expressed via strict rankings~\citep{AS98}. Still, it may produce unfair outcomes as agents who act early have a clear advantage over agents who act later. The obvious way to fix the fairness issue without harming the other two properties is to have the agents act in a uniformly random order. This gives us the well-known {\em random serial dictatorship} (RSD) or {\em random priority} mechanism (see~\cite{AS98} and \cite{BM01}).

In addition to the abstract setting of assignment problems in which agents have {\em ordinal} preferences for the items, we consider two settings which use additional {\em cardinal} information. In the first one (the {\em value setting}), we assume that agents have values for the items and the items of higher value for an agent appear higher in their preference ranking. In the second one (the {\em metric cost setting}), agents and items are assumed to be points in a metric space, and each agent has cost for an item equal to the distance of the corresponding points. Now items of lower cost for an agent are those which appear higher in the preference ranking. In the two settings, we would like to compute matchings with high {\em social welfare} and low {\em social cost}, respectively, defined as the sum of the values in the former case and costs of the agents for their assigned items in the latter.

Even though RSD can neither optimise the social welfare nor the social cost (e.g.\ see~\cite{FFZ14,CFFHT16}), it could be attractive compared to alternatives that do not have its other favourable properties. To explore whether this is the case, we need to be able to assess the outcome of RSD in terms of efficiency. In our two settings, this translates to computing the expected social welfare or the expected social cost of RSD, which in turn requires knowledge of the {\em RSD lottery} matrix. This matrix consists of the probabilities for all agent-item pairs that an agent is assigned to an item by RSD. Unfortunately, computing even a single entry of this matrix is a \#P-complete problem~\citep{ABB13,SS15}. Still, approximations of the expected social welfare and cost would be sufficient to compare RSD with other mechanisms.

So, can we compute ---in reasonable time--- accurate estimates of the expected social welfare and expected social cost of RSD when applied to assignment instances in the value and metric cost settings, respectively? This is the question we study in the current paper.

\subsection{Our Contribution}
We first give a formal argument (in~\Cref{sec:hardness}) explaining how the previous \#P-hardness results of~\cite{ABB13} and~\cite{SS15} for computing the RSD lottery matrix for a given assignment instance in the abstract setting imply the \#P-hardness of computing the expected social welfare and expected social cost of the RSD outcome in the value and metric cost setting, respectively. Specifically, given an assignment instance in the abstract setting, we show how to construct equivalent instances in the value and metric cost setting so that the binary representations of the expected social welfare and expected social cost have polynomial size (in terms of $n$), and furthermore contain the binary representation of the probabilities in the RSD lottery matrix. Then, the existence of a polynomial-time algorithm for computing the expected social welfare/cost in the value/metric cost setting would imply the existence of a polynomial-time algorithm for computing the RSD lottery.

We then consider a simple algorithm which randomly samples a number of agent orderings, applies the serial dictatorship using each of them, and returns the average social welfare (or average social cost) in the computed matchings. In spite of our 
\#P-hardness result, we show that $\Theta\left(\frac{n}{\epsilon^2}\ln{\frac{1}{\delta}}\right)$ samples are sufficient and necessary so that the value returned by the algorithm when applied to assignment instances in the value setting with $n$ agents/items approximates the expected social welfare of RSD within a factor of $1\pm\epsilon$ with probability at least $1-\delta$. These results are presented in~\Cref{sec:approxswelfareofrsd}. To prove the upper bound, we use Bernstein's inequality, which allows us to use a simple bound on the variance that depends on the relation of the expected social welfare of \rsd\ to the optimal social welfare. The lower bound follows by a reverse Chernoff bound.

Unfortunately, in the metric cost setting, the same algorithm needs a much higher number of samples to obtain similar guarantees. Specifically, we show that in order to approximate the expected social cost of \rsd\ within a factor of $1\pm\varepsilon$ with probability at least $1 - \delta$ for all values of parameters $n$, $\varepsilon$, and $\delta$, the number of samples should depend exponentially on either $n$ or $\ln{\frac{1}{\delta}}$. To bypass this barrier, we prove a non-trivial bound on the variance of the social cost of \rsd. This is the most technically interesting among our results and yields that, using only $O\left(\frac{n^3}{\varepsilon^2}\right)$ samples, the algorithm approximates the expected social cost within $1\pm\varepsilon$ with (sufficiently high) constant probability. Then, using an idea from approximate counting and taking the median of values returned by $O\left(\ln{\frac{1}{\delta}}\right)$ executions of the averaging algorithm, we obtain the desired approximation guarantee using at most $O\left(\frac{n^3}{\varepsilon^2}\ln{\frac{1}{\delta}}\right)$ samples in total. These results are presented in \Cref{sec:approxscostofrsd}.

We continue with a discussion on the literature. A quick overview of the sharp concentration bounds we use in our proofs is presented in~\Cref{sec:preliminaries}, together with our notation and definitions. We conclude with~\Cref{sec:discussion}.

\subsection{Further Related Work}
House allocation has been the generic assignment problem; \cite{AS98}, \cite{BM01}, \cite{CM01}, and \cite{SU11} discuss further applications. Besides its simplicity, the SD mechanism has already received considerable attention. For example, in the economic literature \cite{S99} characterized it as the only deterministic assignment mechanism that is strategy-proof, non-bossy and neutral. Several authors (e.g.\ \cite{AS98,ACM+05}) have observed that a solution to the assignment problem is Pareto-optimal if and only if it can be produced by SD with an appropriate agent ordering. In the computer science literature, SD has been studied in the metric cost setting, where it has been proved to be highly inefficient in the worst-case but very efficient under resource augmentation assumptions~\citep{KP93,CFFHT16}. Recently, \cite{CR23} consider the problem of optimizing the agent ordering so that SD yields good results not only in assignment problems but in combinatorial optimization more generally.

The properties of RSD are discussed extensively by \cite{AS98} and \cite{BM01}, where it is also compared to other mechanisms such as the probabilistic serial mechanism and the mechanism of~\cite{HZ79}. The complexity of computing the RSD lottery matrix is studied by~\cite{ABB13} and~\cite{SS15}, who prove that it is \#P-complete; see~\cite{F97} for an introduction to the complexity class \#P. On the positive side, \cite{AM14} present fixed-parameter tractable and polynomial-time algorithms that compute the RSD lottery for restricted assignment instances. In the metric cost model, RSD has been proved to be highly superior to SD, approximating the optimal social cost within a factor that is at most $n$ and at least $n^{0.29}$~\citep{CFFHT16}. We remark that we use some of the results from that paper as well as from the paper by \cite{KP93} in our proofs for the metric cost setting. 

Sampling techniques have found important applications in social choice. Indicative works include their use in deciding the winning alternative according to voting rules or estimates of notions like the distortion~\citep{CMP24} or margin of victory~\citep{BD21}. More related in spirit to our work are papers aiming at estimating the Shapley value in cooperative games~\citep{AK14} or the Banzhaf index in voting~\citep{BMR+10}. 

%% file: preliminaries.tex
An assignment instance consists of $n$ agents and $n$ items. We will use the set $[n]\coloneqq\{1,2, ..., n\}$ to represent both the set of agents and the set of items.

In the {\em abstract setting} usually considered in the literature, each agent $i\in [n]$ has a strict preference ranking of all items. In this paper, we consider two more settings. In the first one, called the {\em value setting}, the agents have {\em values} for the items. For an agent $i\in [n]$ and item $g\in [n]$, we denote by $v_i(g)$ the (non-negative) value agent $i$ has for item $g$. A (perfect) matching $M=(M_1, M_2, ..., M_n)$ is an assignment of the items to the agents so that each item is assigned to one agent, and each agent gets one item. The \textit{social welfare} of a matching is the total value the agents have for the items they are assigned, i.e.\ \(\sw(M) = \sum_{i \in [n]}{v_i(M_i)}\). For an assignment instance $\I$, we denote by $\opt(\I)$ the maximum social welfare among all possible matchings in $\I$.

In the second setting, called the {\em metric cost setting}, the agents have {\em costs} for the items. For an agent $i\in [n]$ and item $g\in [n]$, we denote by $c_i(g)$ the (non-negative) cost agent $i$ has for item $g$. We assume that the agents and items correspond to points in a metric space, and the cost $c_i(g)$ is the distance between the points corresponding to agent $i$ and item $g$. Thus, the costs satisfy the triangle inequality, e.g.\ for agents $i_1$ and $i_2$ and items $g_1$ and $g_2$, the triangle inequality implies that \(c_{i_1}(g_1) \leq c_{i_1}(g_2) + c_{i_2}(g_2) + c_{i_2}(g_1)\). The \textit{social cost} of a matching is the total cost the agents have for their allocated items, i.e.\ \(\soc(M)=\sum_{i \in [n]} {c_i(M_i)}\). For an assignment instance $\I$, we slightly abuse notation and also use $\opt(\I)$ to denote the minimum social cost among all possible matchings in $\I$.

The \textit{serial dictatorship} mechanism (or SD for short) takes an assignment instance $\I$ and an ordering $\pi$ of the agents as input and computes a matching of items and agents as follows. The mechanism considers the agents one by one according to the ordering $\pi$. Whenever an agent is considered, they select their most preferable item that has not been selected by an agent until that point. This will be the agent's highest-ranked item in their preference ranking in the abstract setting, their most valuable item in the value setting, or their least costly item in the metric cost setting. We denote by $\sd(\I,\pi)$ the matching computed by the SD mechanism when applied on instance $\I$ using the agent ordering $\pi$.

The \textit{random serial dictatorship} mechanism (or RSD for short) applies SD using an ordering $\pi$ that has been selected uniformly at random among all agent orderings. The RSD {\em lottery} is an $n\times n$ matrix $P(\I)$ (or simply $P$) in which the entry $P_{i,g}$ denotes the probability that item $g$ is assigned to agent $i$ when RSD is applied on instance $\I$.
We use $\rsd(\I)$ to denote both the expected social welfare in the value setting and the expected social cost in the metric cost setting of the matching returned by RSD, when applied on instance $\I$. For $\epsilon>0$, we say that a quantity $Q$ is an $\epsilon$-approximation of $\rsd(\I)$ if \(|Q-\rsd(\I)|< \epsilon \cdot \rsd(\I)\).
We sometimes use the terms {\em over} and {\em under} $\epsilon$-approximation to refer to a quantity $Q$ satisfying \(Q < (1 + \epsilon) \cdot \rsd(\I)\) and \(Q > (1 - \epsilon) \cdot \rsd(\I)\), respectively.

In the following, we present some inequalities and bounds that we use later on.
The \hyperlink{lem:bernstein_inequality}{Bernstein inequality} is the first one.

\begin{lemma}[Bernstein inequality, e.g.\ see~\cite{DP09}, page 9]\label{lem:bernstein_inequality}
    Let \(X_1, X_2, \dots, X_k\) be independent random variables satisfying \(| X_i | \leq \alpha\) for $i\in [k]$, with mean $0$ and variance $\sigma^2(X_i)$.
    Then
    \[
        \Pr \left[ \left| \sum\limits_{i = 1}^{k} X_i \right| \geq t \right] \leq 2\exp \left( - \frac{3t^2}{6\sum\limits_{i = 1}^{k} \sigma^2 (X_i) + 2\alpha t} \right)
    \]
\end{lemma}

Knowing the mean as well as upper and lower bounds for a random variable, we can use the \hyperlink{lem:bhatia_davis_inequality}{Bhatia-Davis inequality} to bound its variance.

\begin{lemma}[Bhatia-Davis inequality,~\cite{BD00}]\label{lem:bhatia_davis_inequality}
    Consider a random variable $X$ that takes values from the interval $[\alpha,\beta]$ and has expectation $\mu$. Then, its variance is
    \[
        \sigma^2 \leq (\beta - \mu) (\mu - \alpha).
    \]
\end{lemma}

In addition to the \hyperlink{lem:bernstein_inequality}{Bernstein inequality}, we will use the \hyperlink{lem:chebyshev-cantelli}{Chebyshev-Cantelli inequality} and the \hyperlink{lem:chernoff}{Chernoff bound} for proving (one-sided) concentration bounds.

\begin{lemma}[Chebyshev-Cantelli inequality, e.g.\ see~\cite{MR95}, page 64]\label{lem:chebyshev-cantelli}
    Let \(X\) be a random variable with expectation \(\mu\) and variance \(\sigma^2\).
    Then, for \(t > 0\), it holds
    \[
        \Pr\left[X - \mu \geq t\sigma\right] \leq \frac{1}{1 + t^2}\, .
    \]
\end{lemma}

\begin{lemma}[Chernoff bound, e.g. see~\cite{MR95}, page 68]\label{lem:chernoff}
    Let \(X_1, X_2, \dots, X_k\) be independent random variables taking values in \(\{0, 1\}\) and \(X = \sum_{i=1}^k{X_i}\) be their sum with expectation \(\mathbb{E}[X] = \mu\).
    Then, for \(\eta>0\), it holds
    \[
        \Pr\left( X \geq (1 + \eta) \mu \right) \leq \left( \frac{e^{\eta}}{(1 + \eta)^{1 + \eta}} \right)^{\mu}.
    \]
\end{lemma}

Moreover, we will use an anti-concentration bound, also known as the \hyperlink{lem:reverse-chernoff}{reverse Chernoff bound}.

\begin{lemma}[Reverse Chernoff bound,~\cite{KY15}]\label{lem:reverse-chernoff}
    Let $X_1, X_2, \dots, X_k$ be independent and identically distributed Bernoulli random variables with expectation $p\in (0,1/2]$.
    Then, for every $\eta\in (0,1/2]$ so that $\eta^2pk\geq 3$, it holds
    \[
        \Pr\left[\frac{1}{k}\sum_{i=1}^k{X_i}\geq (1+\eta)p\right] \geq \exp\left(-9\eta^2pk\right).
    \]
\end{lemma}


%% file: hardness.tex
Before presenting our estimation results, let us explain how the \#P-hardness of computing the RSD lottery (\citet{ABB13,SS15}) for a given assignment instance implies the \#P-hardness of computing the expected social welfare or the expected social cost in the value or metric cost setting, respectively.

The above-mentioned papers deal with assignment instances in abstract form. For $j\in [n]$, let $r_i(j)$ denote the $j$-th most preferred item of agent $i\in [n]$. Given such an instance in abstract form, we show how to construct equivalent instances in the value and metric cost settings, in the sense that the outcome of RSD applied to the three instances, and consequently their RSD lotteries, coincide.

Given the preference orderings $r_i$ of each agent $i\in [n]$, we define consistent agent values $v_i$ and metric costs $c_i$ as follows:
\begin{itemize}
    \item In the value setting, we define the value of agent $i$ for item $r_i(j)$ to be $2^{\left(in-j\right)\left\lceil \log{\left(n!+1\right)}\right\rceil}$. 
    \item In the metric cost setting, we define the cost of agent $i$ for item $r_i(j)$ to be $2^{n^2\left\lceil \log{\left(n!+1\right)}\right\rceil}+2^{\left((i-1)n+j-1\right)\left\lceil \log{\left(n!+1\right)}\right\rceil}$. Notice that all agent costs differ by less than a multiplicative factor of $2$ and, thus, they define a metric.
\end{itemize}
Also, notice that the number of bits in the representation of values and costs is at most $O(n^3\log{n})$. Hence, the construction of the instances in the value and metric cost setting takes only polynomial time.

Let $P$ be the RSD lottery and $L$ be the $n\times n$ matrix with entry $L_{ij}$ being the number of different agent orderings $\pi$ so that $\sd(\I,\pi)$ assigns item $r_i(j)$ to agent $i$. Clearly, $L_{ij}=n!\cdot P_{i,r_i(j)}$ for every $i,j\in [n]$.

In the value setting, we have 
\begin{align*}
\sum_{j\in [n]}{P_{ij}\cdot v_i(j)} &= \sum_{j\in [n]}{P_{i,r_i(j)}\cdot v_i(r_i(j))}=\frac{1}{n!}\cdot \sum_{j\in [n]}{L_{ij}\cdot v_i(r_i(j))}
\end{align*}
for agent $i\in [n]$. Thus, we have
\begin{align*}
n!\cdot \E[\sw(\sd(\I,\pi))] &= n!\cdot \sum_{i\in [n]}{\sum_{j\in [n]}{P_{ij}\cdot v_i(j)}}=\sum_{i\in [n]}{\sum_{i\in [n]}{L_{ij}\cdot v_i(r_i(j))}},
\end{align*}
i.e.\ the quantity $n!\cdot \E[\sw(\sd(\I,\pi))]$ is a non-negative integer. Now, recall that $v_i(r_i(j))=2^{(in-j)\left\lceil \log{(n!+1)}\right\rceil}$, and thus the binary representation of the integer $L_{ij}\cdot v_i(r_i(j))$ has the binary representation of $L_{ij}$ in bit positions\footnote{We number the bit positions by assuming that the least significant bit is at position $0$.} from $(in-j)\left\lceil \log{(n!+1)}\right\rceil$ to $(in-j+1)\left\lceil \log{(n!+1)}\right\rceil-1$ and $0$s everywhere else. Notice that these bit positions are disjoint for different pairs of $i$ and $j$ in $[n]$; indeed, $\left\lceil \log{(n!+1)}\right\rceil$ bits are enough to encode $L_{ij}$, which can take (integer) values between $0$ and $n!$. Thus, the binary representation of $n!\cdot \E[\sw(\sd(\I,\pi))]$ has the binary representation of $L_{ij}$ in bit positions from $(in-j)\left\lceil \log{(n!+1)}\right\rceil$ to $(in-j+1)\left\lceil \log{(n!+1)}\right\rceil-1$ for every $i,j\in [n]$. 

Similarly, in the metric cost setting, we have
\begin{align*}
\sum_{j\in [n]}{P_{ij}\cdot c_i(j)}
&=\frac{1}{n!}\cdot \sum_{i\in [n]}{L_{ij}\cdot c_i(r_i(j))},
\end{align*}
and, hence,
\begin{align*}
n!\cdot \E[\soc(\sd(\I,\pi))] &= n!\cdot \sum_{i\in [n]}{\sum_{j\in [n]}{P_{ij}\cdot c_i(j)}}=\sum_{i\in [n]}{\sum_{i\in [n]}{L_{ij}\cdot c_i(r_i(j))}},
\end{align*}
i.e.\ the quantity $n!\cdot \E[\soc(\sd(\I,\pi))]$ is again a non-negative integer. Now, recall that $c_i(r_i(j))=2^{n^2\left\lceil \log{(n!+1)}\right\rceil}+2^{(i-1)n+j-1)\left\lceil \log{(n!+1)}\right\rceil}$, and thus the binary representation of the integer $L_{ij}\cdot c_i(r_i(j))$ has the binary representation of $L_{ij}$ in bit positions from $((i-1)n+j-1)\left\lceil \log{(n!+1)}\right\rceil$ to $((i-1)n+j)\left\lceil \log{(n!+1)}\right\rceil-1$, $1$ in bit position $n^2\left\lceil\log{(n!+1)}\right\rceil$, and $0$s everywhere else. Thus, the binary representation of $n!\cdot \E[\soc(\sd(\I,\pi))]$ has the binary representation of $n^2$ in the $\left\lceil \log(n^2)\right\rceil$ bit positions from $n^2\left\lceil\log{(n!+1)}\right\rceil$ to $ n^2\left\lceil\log{(n!+1)}\right\rceil+\left\lceil \log(n^2)\right\rceil-1$, and the binary representation of $L_{ij}$ in the bit positions from $((i-1)n+j-1)\left\lceil \log{(n!+1)}\right\rceil$ to $((i-1)n+j)\left\lceil \log{(n!+1)}\right\rceil-1$ for every $i,j\in [n]$. 

From the discussion above we conclude that any polynomial-time algorithm that computes the expected social welfare in the value setting or the expected social cost in the metric cost setting can be used to compute the entries of the matrix $L$ in polynomial time, and thus the RSD lottery of the assignment instance. The following statement summarises the discussion above.

\begin{theorem}
    Given an assignment instance $\I$ in the value or metric cost setting, computing the expected social welfare or expected social cost of the outcome of RSD when applied on $\I$ is \#P-hard.
\end{theorem}

%% file: approxsocialwelfare.tex
We now consider a simple algorithm (\Cref{alg:approx_sw_of_rsd}) that estimates (approximates) the expected social welfare of \rsd. Given an assignment instance $\I$ in the value setting, \Cref{alg:approx_sw_of_rsd} samples $k$ agent orderings (uniformly at random and with replacement), computes $k$ matchings by applying SD on $\I$ using each of the sampled agent orderings, and returns the average of the social welfare of these matchings as output. We will present upper and lower bounds on $k$ so that \Cref{alg:approx_sw_of_rsd} returns an $\epsilon$-approximation of $\rsd(\I)$ with probability at least $1-\delta$. 

\begin{algorithm}[h]
    \caption{Approximating the expected social welfare of Random Serial Dictatorship}
    \label{alg:approx_sw_of_rsd}
    \begin{algorithmic}[1]
        \Input An assignment instance $\I$ with $n$ agents/items and an integer \(k\geq 1\)
        \Output A non-negative number
        \State Select independently $k$ uniformly random orderings \(\pi^1, \pi^2, \dots, \pi^k\)
        \State\Return \(\tfrac{1}{k} \sum\limits_{i = 1}^k{\sw(\sd(\I,\pi^i))} \)
    \end{algorithmic}
\end{algorithm}

We start with a simple lemma that relates the maximum social welfare and the expected social welfare of the RSD outcome. 
This will come in handy in the proof of \Cref{thm:sw_rsd_upper_bound} below.

\begin{lemma}\label{lem:bound_approx_sw_of_rsd_with_max_sw}
    For every assignment instance $\I$ with $n$ agents/items in the value setting, it holds \(\opt(\I) \leq n\cdot \rsd(\I)\).
\end{lemma}

\begin{proof}
Let \(M=(M_1, M_2, \dots, M_n)\) be the matching of maximum social welfare in the assignment instance $\I$. For \(i\in [n]\), denote by $g_i$ the item of maximum value for agent $i$. Notice that the RSD mechanism allocates item $g_i$ to agent $i$ with a probability of at least \(1/n\). Then,
\begin{align*}
    \rsd(\I) &\geq \frac{1}{n}\cdot \sum_{i=1}^n{v_i(g_i)}\geq \frac{1}{n}\cdot \sum_{i=1}^n{v_i(M_i)}=\opt(\I)/n,
\end{align*}
as desired.
\end{proof}

We are now ready to prove our upper bound. The key ideas in the proof are the use of the Bhatia-Davis inequality (\Cref{lem:bhatia_davis_inequality}) to bound the variance of the social welfare of the matching returned by RSD and the application of Bernstein inequality (\Cref{lem:bernstein_inequality}).\footnote{Readers familiar with probabilistic analysis may wonder why we do not use the simpler Hoeffding inequality to prove (a statement similar to) \Cref{thm:sw_rsd_upper_bound}. We present such an analysis in \Cref{sec:app:A}, which yields a weaker bound of $O\left(\frac{n^2}{\varepsilon^2}\ln{\frac{1}{\delta}}\right)$ on $k$. There, we also argue that the main reason for this weaker result is the lack of information about the variance; something that the Bernstein inequality exploits.}

\begin{theorem}\label{thm:sw_rsd_upper_bound}
    Let \(n\geq 1\) be an integer and \(\delta,\epsilon\in (0,1]\). For \(k \geq \tfrac{8n}{3 \epsilon^2} \ln \tfrac{2}{\delta}\), the output of \Cref{alg:approx_sw_of_rsd}, when applied on the assignment instance $\I$ with $n$ agents/items in the value setting, is an $\epsilon$-approximation to the expected social welfare of the RSD mechanism with probability at least $1-\delta$.
\end{theorem}

\begin{proof}
    For $i\in [k]$ and a uniformly random ordering \(\pi^i\) of the agents in $[n]$, notice that the random variable \(\sw(\sd(\I,\pi^i))\) takes values in \([0,\opt(\I)]\) and has expectation $\rsd(\I)$. By the Bhatia-Davis inequality (\Cref{lem:bhatia_davis_inequality}), we have that the variance of the random variable $\sw(\sd(\I,\pi^i))$ is
    \begin{align}\nonumber
        \sigma^2(\sw(\sd(\I,\pi^i))) &\leq (\opt(\I) - \rsd(\I))\cdot \rsd(\I)\\\label{eq:bound-variance-of-SD}
        &\leq \opt(\I) \cdot \rsd(\I).
    \end{align}
    Now, for \(i\in [k]\), define the random variable $Z_i$ as \(Z_i\coloneqq\sw(\sd(\I,\pi^i)) - \rsd(\I)\). We have that the probability that the output of \Cref{alg:approx_sw_of_rsd} is not an $\epsilon$-approximation of $\rsd$ is
    \begin{align}\label{eq:bad-approximation}
        \Pr\left[\left|\frac{1}{k}\sum_{i=1}^k{\sw(\sd(\I,\pi^i))} - \rsd(\I) \right| \geq \epsilon \cdot \rsd(\I)\right]
        &=\Pr \left[\left|\sum_{i=1}^k{Z_i}\right| \geq \epsilon k \cdot \rsd(\I)\right].
    \end{align}
    To complete the proof, we will bound the RHS of \Cref{eq:bad-approximation}. By the definition of the random variable $Z_i$, its variance is equal to the variance of the random variable $\sw(\sd(\I,\pi^i))$, i.e.\ by \Cref{eq:bound-variance-of-SD}, it holds that
    \[
        \sigma^2(Z_i) \leq \opt(\I) \cdot \rsd(\I).
    \]
    Furthermore, \(Z_i\) takes values in \([-\opt(\I), \opt(\I)]\) and has expectation $0$.

    We now apply Bernstein inequality (\Cref{lem:bernstein_inequality}) for the random variable \(\sum_{i=1}^k{Z_i}\) using \(t=\epsilon k\cdot \rsd(\I)\) and \(\alpha=\opt(\I)\). By the discussion above, we have \(\sum_{i=1}^k{\sigma^2(Z_i)}\leq k\cdot \opt(\I) \cdot \rsd(\I)\). Thus,
    \begin{align}\nonumber
        \Pr \left[\left|\sum_{i=1}^k{Z_i}\right| \geq \epsilon k \cdot \rsd(\I)\right]
        &\leq 2 \exp \left(-\frac{3\epsilon^2 k^2 \cdot \rsd(\I)^2}{6k \cdot \opt(\I) \cdot \rsd(\I) + 2\epsilon k \cdot \opt(\I) \cdot \rsd(\I)}\right)\\\label{eq:applying-Bernstein-ineq}
        &\leq 2 \exp \left(-\frac{3\epsilon^2 \cdot k\cdot \rsd(\I)}{8\cdot \opt(\I)}\right) \leq 2 \exp \left(-\frac{3\epsilon^2 \cdot k}{8n}\right).
    \end{align}
    The second inequality follows since $\epsilon\leq 1$ and the third one by \Cref{lem:bound_approx_sw_of_rsd_with_max_sw}. By \Cref{eq:bad-approximation} and \Cref{eq:applying-Bernstein-ineq}, we conclude that for $k\geq \frac{8n}{3\epsilon^2}\cdot \ln\frac{2}{\delta}$, we have
    \[
        \Pr \left[\left|\frac{1}{k}\sum_{i=1}^k{\sw(\sd(\I,\pi^i))}-\rsd(\I)\right| \geq \epsilon \cdot \rsd(\I)\right] \leq \delta,
    \]
    as desired.
\end{proof}

We now prove our lower bound for the value setting, by applying the reverse Chernoff bound (\Cref{lem:reverse-chernoff}). 

\begin{theorem}\label{thm:sw_rsd_lower_bound}
    Let $n\geq 2$ be an integer, \(\epsilon\in (0,1]\), \(\delta\in (0,e^{-27})\), and $k$ be such that \(\tfrac{3n}{\epsilon^2} \leq k < \tfrac{n}{9 \epsilon^2} \ln \tfrac{1}{\delta}\). Then, there exists an assignment instance $\I$ with $n$ agents/items in the value setting, so that the output of \Cref{alg:approx_sw_of_rsd} when applied on $\I$ is an $\epsilon$-approximation to the expected social welfare of the RSD mechanism with probability smaller than $1-\delta$.
\end{theorem}

\begin{proof}
    Let $\I$ be the instance in which agent $1$ has value $1$ for item $1$ and $0$ for any other items. All other agents have a valuation of $0$ for all items. All ties regarding the item an agent picks under SD are resolved in favour of the minimum-index item. Thus, the serial dictatorship returns a matching of social welfare $1$ when applied with an ordering that has agent $1$ first and social welfare $0$ otherwise. So, for a uniformly random ordering $\pi$ of the agents, \(\sw(\sd(I,\pi))\) is a Bernoulli random variable with expectation $1/n$. Thus, \(\rsd(\I)=1/n\). We will use the reverse Chernoff bound (\Cref{lem:reverse-chernoff}) to bound the probability that \Cref{alg:approx_sw_of_rsd} computes an $\epsilon$-approximation of $\rsd(\I)$ from below.

    We apply \Cref{lem:reverse-chernoff} to the random variables \(X_1, X_2, \dots, X_k\) denoting the $k$ independent copies of the random variable \(\sw(\sd(I,\pi))\) used by \Cref{alg:approx_sw_of_rsd}. Notice that we have \(p=1/n\), meaning that the lower bound on $k$ in the statement of the theorem guarantees that \(\epsilon^2pk\geq 3\), and hence the conditions of \Cref{lem:reverse-chernoff} are satisfied. We obtain that the probability that the quantity returned by \Cref{alg:approx_sw_of_rsd} is not an $\epsilon$-approximation is
    \begin{align*}
        &\Pr\left[\left|\frac{1}{k}\sum_{i=1}^k{\sw(\sd(\I,\pi^i))}-\rsd(\I)\right| \geq \epsilon \cdot \rsd(\I)\right]\\
        &\geq \Pr\left[\frac{1}{k}\sum_{i=1}^k{X_i}\geq (1+\epsilon)\rsd(\I)\right] \geq \exp\left(-\frac{9\epsilon^2k}{n}\right)
        >\delta,
    \end{align*}
    implying that the probability that the output of \Cref{alg:approx_sw_of_rsd} is an $\epsilon$-approximation to the expected social welfare of random serial dictatorship is less than $1-\delta$.
\end{proof}

%% file: approxsocialcost.tex
To approximate the expected social cost of RSD in the metric cost setting, we can modify \Cref{alg:approx_sw_of_rsd} by changing $\sw$ with $\soc$ in Line 2, i.e.\ the algorithm now returns the average social cost of the $k$ matchings returned by executing SD with each of the $k$ random agent orderings. We will refer to this modification as \Cref{alg:approx_sw_of_rsd} as well.

Unfortunately, as we show in \Cref{subsec:lower-cost}, to return an $\varepsilon$-approximation with probability at least $1-\delta$ for all assignment instances and all values of parameters, \Cref{alg:approx_sw_of_rsd} must use a value for $k$ that depends exponentially on either $n$ or $\ln{\tfrac{1}{\delta}}$.\footnote{In \Cref{sec:app:B}, we present such upper bounds for \Cref{alg:approx_sw_of_rsd} by adapting ideas from \Cref{sec:approxswelfareofrsd} and combining them with our new bounds on the variance of \rsd.} To bypass this issue, we use a technique from the literature on approximate counting (see \cite[Chapter 28]{V01}) by executing \Cref{alg:approx_sw_of_rsd} several times and taking the median value returned in these executions. This is implemented in \Cref{alg:approx_sc_of_rsd} below.

\begin{algorithm}[ht]
    \caption{Approximating the expected social cost of Random Serial Dictatorship}
    \label{alg:approx_sc_of_rsd}
    \begin{algorithmic}[1]
        \Input An assignment instance $\I$ with $n$ agents/items and  integers \(k, \lambda \geq 1\)
        \Output A non-negative number
        \For{\(j\gets 1,2,\dots\lambda\)}
            \State \(\xi_j\gets \text{\hyperlink{alg:approx_sw_of_rsd}{Algorithm1}}(\I, k)\) \label{alg2:line2}
        \EndFor
        \State\Return \(\median(\boldsymbol{\xi}) \)
    \end{algorithmic}
\end{algorithm}

For the analysis of \Cref{alg:approx_sc_of_rsd}, we will need upper bounds on the variance of \rsd. Unfortunately, while \(0\) and \(\opt(\I)\) are natural bounds on the social welfare of a matching returned by RSD when applied on an assignment instance $\I$ in the value setting, the corresponding bounds for the social cost in the metric cost setting are much further apart, and the Bhatia-Davis inequality is not useful anymore in bounding the variance. Instead, we prove a new bound on the variance of the social cost returned by \rsd, which we present in \Cref{subsec:variance}. Finally, in \Cref{subsec:upper-cost}, we prove bounds on the parameters $k$ and $\lambda$ used by \Cref{alg:approx_sc_of_rsd} so that it computes an $\varepsilon$-approximation of $\rsd$.

\subsection{A Lower Bound for \Cref{alg:approx_sw_of_rsd}}\label{subsec:lower-cost}
We begin our study of the metric cost setting by showing an exponential lower bound on the number of samples needed by \Cref{alg:approx_sw_of_rsd}.

\begin{theorem}\label{thm:rsd_lower_bound_sc}
If \Cref{alg:approx_sw_of_rsd} returns an $\epsilon$-approximation to the expected social cost of the \rsd\ mechanism with probability at least $1-\delta$ on input any assignment instance in the metric cost setting with $n$ agents/items and for every $\delta,\epsilon\in (0,1)$, then $k$ should depend exponentially on either $n$ or $\ln{\frac{1}{\delta}}$. \end{theorem}

\begin{proof}
    We use a family of assignment instances that are very similar to the worst-case instances used by~\cite{CFFHT16} (see also~\citet{KP93}) to prove lower bounds on the performance of serial dictatorship in the metric cost setting. 

For $n\geq 1$, the assignment instance $\I_n$ is defined as follows. There are $n$ agents at locations \(1, 2, 4, \dots, 2^{n-1}\) and $n$ items at locations \(-1, 2, 4, \dots, 2^{n-1}\) on the real line. Notice that the assignment which matches the agent at location $1$ to the item at location $-1$ and, for $i=1, ..., n-1$, the agent at location $2^i$ to the item at the same location has a social cost of $2$. Thus, $\opt(\I_n)\leq 2$. 

Now, consider the agent ordering $\pi^*_n=\langle 1,2,...,n\rangle$ and observe that the execution of serial dictatorship on instance $\I_n$ using this agent ordering returns the assignment in which, for $i=1, 2, ..., n-1$, the agent at location $2^{i-1}$ is matched to the item at location $2^i$, and the agent at location $2^{n-1}$ is matched to the item at location $-1$. Thus, $\soc(\sd(\I_n,\pi^*_n))=\sum_{i=1}^{n-1}{2^{i-1}}+2^{n-1}+1=2^n$.

For $n\geq 1$, consider the instance $\I_n$ of the above family with $n$ agents/items. Let $\delta=\frac{1}{2n!}$ and $\epsilon\in (1/2,1)$, and assume, for the sake of contradiction, that $k\leq \frac{2^n}{4n}$. The probability that \Cref{alg:approx_sw_of_rsd} does not return an $\epsilon$-approximation on input instance $\I_n$ is
\begin{align}\nonumber
    &\Pr\left[\left|\frac{1}{k}\sum_{i=1}^k{\soc(\sd(\I_n,\pi^i))-\rsd(\I_n)}\right|\geq \epsilon\cdot \rsd(\I_n)\right]\\\nonumber
    &\geq \Pr\left[\frac{1}{k}\sum_{i=1}^k{\soc(\sd(\I_n,\pi^i))}\geq 2\cdot \rsd(\I_n)\right]\\\label{eq:prob_ineq}
    &\geq \Pr\left[\sum_{i=1}^k{\soc(\sd(\I_n,\pi^i)) \geq 4kn}\right]
    \geq \Pr\left[\sum_{i=1}^k{\soc(\sd(\I_n,\pi^i)) \geq 2^n}\right]\,.
\end{align}
The first inequality follows since $\epsilon<1$, the second one from the result of~\cite{CFFHT16}, stating that $\rsd(\I_n)\leq n\cdot \opt(\I_n)\leq 2n$, and the third inequality follows since $k\leq\frac{2^n}{4n}$.

Now, recall that $\soc(\sd(\I_n,\pi^*_n))\geq 2^n$; thus, the probability in the last line of derivation (\ref{eq:prob_ineq}) is lower-bounded by the probability that the ordering $\pi^*_n$ is selected as one of the $k$ orderings used by \Cref{alg:approx_sw_of_rsd} which in turn is at least $1/n!>\delta$. Thus,
\begin{align}\nonumber
    \Pr\left[\left|\frac{1}{k}\sum_{i=1}^k{\soc(\sd(\I_n,\pi^i))-\rsd(\I_n)}\right|\geq\epsilon\cdot \rsd(\I_n)\right] &>\delta,
\end{align}
which implies that \Cref{alg:approx_sw_of_rsd} returns an $\epsilon$-approximation of $\rsd(\I_n)$ with probability at least $1-\delta$ only when $k>\frac{2^n}{4n}$. In this case, $k$ depends exponentially on either $n$ or $\ln{1/\delta}<n^2$. The theorem follows.
\end{proof}

\subsection{Bounding the Variance of Social Cost}\label{subsec:variance}
We will shortly turn our attention to \Cref{alg:approx_sc_of_rsd}. For its analysis, we will prove an upper bound on the variance of the social cost of \rsd, or more precisely, on the expectation of the square of its social cost; we do so in the following lemma.

\begin{lemma}\label{lem:exp-sc-square}
    For every assignment instance \(\I\) with \(n\) agents/items in the metric cost setting, it holds that
    \[
        \E [\soc(\sd(\I, \pi))^2] \leq n^3 \cdot OPT(\I)^2\, ,
    \]
    where \(\pi\) is a uniformly random ordering of the agents.
\end{lemma}

\begin{proof}
    We will prove the statement using induction on $n$.
    Observe that the statement holds trivially if \(n = 1\) since there is a single perfect matching in this case.
    Assuming the statement holds for all assignment instances with \(n - 1\) agents/items in the metric cost setting, we show that this is true for instances with \(n\) agents/items as well.
    
    We make use of some additional notation throughout the proof.
    For any \(i \in [n]\), we denote by $r_i$ the item that agent $i$ prefers the most (breaking ties arbitrarily). Also, for any $i\in [n]$, given an assignment instance \(\I\) with \(n\) agents/items, we denote by \(\I_{-i}\) the assignment instance obtained by \(\I\) after removing agent \(i\) and item \(r_i\).
    We also use \(\pi_{-i}\) to denote a uniformly random ordering of the agents in \([n] \setminus \{i\}\).
    We can view the execution of \rsd\ as a uniformly random selection of the first agent \(i\) who picks their most preferred item \(r_i\) followed by running the \rsd\ mechanism on the reduced instance \(\I_{-i}\).
    Therefore, we have
    \begin{align}\nonumber 
        \E \left[ \soc(\sd (\I, \pi))^2 \right] 
        &= \frac{1}{n} \sum\limits_{i = 1}^{n} \E \left[ \left( c_i (r_i) + \soc (\sd (\I_{-i}, \pi_{-i})) \right)^2 \right] \\\nonumber
        &= \frac{1}{n} \sum\limits_{i = 1}^{n} c_i (r_i)^2 + \frac{2}{n} \sum\limits_{i = 1}^{n} c_i(r_i) \cdot \E \left[ \soc (\sd (\I_{-i}, \pi_{-i})) \right]\\\label{eq:exp-sc-1}
            &\quad + \frac{1}{n} \sum\limits_{i = 1}^{n} \E \left[ \soc \left( \sd (\I_{-i}, \pi_{-i}) \right)^2 \right] \, .
    \end{align}

    We proceed by presenting two claims which will be useful to upper-bound the two final terms of the RHS in \Cref{eq:exp-sc-1}.
    \begin{claim}\label{cl:sum-of-best-bound}
    For every assignment instance \(\I\) with \(n\) agents/items in the metric cost setting, it holds that
    \(\sum_{i = 1}^{n} c_i(r_i)\leq \opt(\I)\) 
    and \(\sum_{i = 1}^{n} c_i \left( r_i \right)^2 \leq \opt(\I)^2\).
    \end{claim}
    
    \begin{proof}
        Consider any perfect matching \(M\) for the assignment instance \(\I\).
        Clearly, for every \(i \in [n]\), the cost of agent \(i\) for the item she is matched to in \(M\) is at least \(c_i(r_i)\).
        Thus, \(\opt(\I) \geq \sum_{i = 1}^{n} c_i(r_i)\) and
        \(\opt(\I)^2 \geq \left( \sum_{i = 1}^{n} c_i(r_i) \right)^2 \geq \sum_{i = 1}^{n} c_i\left( r_i \right)^2\), as desired. 
    \end{proof}
    
    The next claim follows from~\cite{CFFHT16}, who proved that the expected social cost of the outcome of RSD on any assignment instance with \(n\) agents/items in the metric cost setting is at most \(n\) times the optimal social cost.
    \begin{claim}\label{cl:exp-soc-cost-bound}
        For every assignment instance \(\I\) with \(n\) agents/items in the metric cost setting and every agent \(i\in [n]\), it holds that
        \(\E \left[ \soc (\I_{-i}, \pi_{-i}) \right] \leq (n - 1) \cdot \opt(\I_{-i})\).
    \end{claim}

    By the induction hypothesis, we have
    \begin{equation}
        \E \left[ \soc \left( \sd (\I_{-i}, \pi_{-i}) \right)^2 \right] \leq (n - 1)^3 \cdot \opt (\I_{-i})^2 \, .
        \label{eq:exp-sc-2}
    \end{equation}
    Using \Cref{cl:sum-of-best-bound} and \Cref{cl:exp-soc-cost-bound} as well as \Cref{eq:exp-sc-2}, \Cref{eq:exp-sc-1} yields
    \begin{align}\nonumber
        \E \left[ \soc \left( \sd (\I, \pi) \right)^2 \right]
            &\leq \frac{1}{n} \cdot \opt (\I)^2 + \left( 2 - \frac{2}{n} \right) \sum\limits_{i = 1}^{n} c_i (r_i) \cdot \opt(\I_{-i})\\\label{eq:exp-sc-3}
            &\quad + \frac{(n-1)^3}{n} \sum\limits_{i = 1}^{n} \opt(\I_{-i})^2 \, .
    \end{align}
    We now use a claim that has also been used by~\cite{CFFHT16}.
    The proof has been included here for the sake of completeness.
    \begin{claim}\label{cl:soc-cost-opt-bound}
        For every assignment instance \(\I\) with \(n\) agents/items in the metric cost setting and every agent \(i\in [n]\), it holds that
        \(\opt(\I_{-i}) \leq \opt(\I) + c_i(r_i).\)
    \end{claim}
    
    \begin{proof}
        Let \(M\) be a matching of minimum social cost on instance \(\I\).
        If agent \(i\) is matched to item \(r_i\) in \(M\), then the restriction of \(M\) that does not include the pair \(\left(i, r_i\right)\) is a matching of \(\I_{-i}\) of social cost at most \(\opt(\I)\), and the statement follows.
        Assume now that agent \(i\) is matched to some item \(g\) different from \(r_i\) in \(M\), while some agent \(j \neq i\) is matched to item \(r_i\).
        The set of agent-item pairs consisting of pair \((j, g)\) and the restriction of \(M\) not including the pairs \((i, g)\) and \((j, r_i)\) is a matching for instance \(\I_{-i}\) of social cost
        \[
            \opt(\I) - c_i(g) - c_j(r_i) + c_j(g) \leq \opt(\I) + c_i(r_i) \, .
        \]
        The inequality follows from applying the triangle inequality which states that
        \(c_j(g) \leq c_j(r_i) + c_i(r_i) + c_i(g)\).
    \end{proof}

    \noindent
    Using \Cref{cl:sum-of-best-bound} and \Cref{cl:soc-cost-opt-bound}, the sum in the second term of the RHS of \Cref{eq:exp-sc-3} becomes
    \begin{align}\nonumber
        \sum\limits_{i = 1}^{n} c_i(r_i) \cdot \opt(\I_{-i})
        &\leq \sum\limits_{i = 1}^{n} c_i(r_i) \cdot \left( \opt(\I) + c_i(r_i) \right)\\\label{eq:exp-sc-4}
        &= \opt(\I) \cdot \sum\limits_{i = 1}^{n} c_i(r_i) + \sum\limits_{i = 1}^{n} c_i(r_i)^2  
        \leq 2 \cdot \opt(\I)^2 \, .
    \end{align}
    Similarly, making use of \Cref{cl:sum-of-best-bound} and \Cref{cl:soc-cost-opt-bound} once more, the sum in the third term of the RHS of \Cref{eq:exp-sc-3} results in
    \begin{align}\nonumber
        \sum\limits_{i = 1}^{n} \opt(\I_{-i})^2 &\leq \sum\limits_{i = 1}^{n} \left( \opt(\I) + c_i(r_i) \right)^2\\\nonumber
        &= n \cdot \opt(\I)^2 + 2 \opt(\I) \cdot \sum\limits_{i = 1}^{n} c_i(r_i) + \sum\limits_{i = 1}^{n} c_i(r_i)^2\\\label{eq:exp-sc-5}
        &\leq (n + 3) \cdot \opt(\I)^2\, .
    \end{align}
    Finally, using \Cref{eq:exp-sc-4} and \Cref{eq:exp-sc-5}, \Cref{eq:exp-sc-3} yields
    \begin{align*}
        \E \left[ \soc \left(\sd (\I, \pi)\right)^2 \right]
        &\leq \left( \frac{1}{n} + 2 \left( 2 - \frac{2}{n} \right) + \frac{(n - 1)^3}{n} (n + 3) \right) \cdot \opt(\I)^2\\
        &= \left( n^3 - \frac{6}{n} (n - 1)^2\right) \opt(\I)^2 \leq n^3 \cdot \opt(\I)^2 \, ,
    \end{align*}
    as desired.
\end{proof}

\subsection{The Upper Bound for \Cref{alg:approx_sc_of_rsd}}\label{subsec:upper-cost}
We are now ready to present bounds on the parameters $k$ and $\lambda$ so that \Cref{alg:approx_sc_of_rsd} computes an $\varepsilon$-approximation of \rsd\ with probability at least $1-\delta$. This requires sampling only $O\left(\frac{n^3}{\varepsilon^2}\ln{\frac{1}{\delta}}\right)$ agent orderings and running \rsd\ according to them. The proof has two parts. First, in \Cref{lem:sc_rsd_single_instance}, we exploit our bound on the variance of the social cost of \rsd\ from \Cref{lem:exp-sc-square} to prove that the probability that an execution of \Cref{alg:approx_sw_of_rsd} does not return an over or an under $\varepsilon$-approximation is at most $1/4$. This is enough to conclude (using a Chernoff bound argument in the proof of \Cref{thm:sc_rsd_upper_bound}) that the median of the social cost values returned by all executions of \rsd\ provides the desired $\varepsilon$-approximation with high probability.

\begin{lemma}\label{lem:sc_rsd_single_instance}
    Let \(n \geq 1\) be an integer and \(\varepsilon \in (0, 1]\).
    For \(k \geq \frac{3n^3}{\epsilon^2}\), the probability that the output of \Cref{alg:approx_sw_of_rsd}, when applied on the assignment instance \(\I\) with \(n\) agents/items in the metric cost setting, is not an over (respectively, not an under) \(\varepsilon\)-approximation to the expected social cost of RSD is at most \(\tfrac{1}{4}\).
\end{lemma}

\begin{proof}
    By \Cref{lem:exp-sc-square}, for \(i\in [k]\), the random variable \(\soc(\sd(\I,\pi^i))\) has variance at most \(n^3\cdot \opt(\I)^2\). Thus, the random variable \(\frac{1}{k}\sum_{i=1}^k{\soc(\sd(\I,\pi^i))}\) has expectation \(\rsd(\I)\) and variance \(\sigma^2\leq \frac{n^3}{k}\cdot \opt(\I)^2\), since the random orderings $\pi^i$ are independent for $i\in [k]$. We will now apply the Chebyshev-Cantelli's inequality (\Cref{lem:chebyshev-cantelli}) to the random variable \(\frac{1}{k}\sum_{i=1}^k{\soc(\sd(\I,\pi^i))}\) with \(t=\frac{\sqrt{k} \cdot \varepsilon \cdot \rsd(\I)}{\sigma}\). Notice that using the condition on \(k\) as well as the fact that \(\opt(\I)\leq \rsd(\I)\), we get
    \[ t^2 = \frac{k\cdot \varepsilon^2 \cdot \rsd(\I)^2}{\sigma^2}\geq \frac{k\cdot \varepsilon^2\cdot \rsd(\I)}{n^3\cdot \opt(\I)}\geq \frac{k\cdot \varepsilon^2}{n^3}\geq 3.\]
    \Cref{lem:chebyshev-cantelli} then yields
    \[
        \Pr\left[\frac{1}{k}\sum_{i=1}^k{\soc(\sd(\I,\pi^i))} \geq (1 + \varepsilon) \cdot \rsd(\I) \right]
        \leq \frac{1}{4}\, .
    \]

Hence, the probability that the output of \Cref{alg:approx_sw_of_rsd} is an over $\varepsilon$-approximation to $\rsd(\I)$ is at least $\tfrac{1}{4}$, as desired. The proof for it being an under \(\varepsilon\)-approximation follows along the same lines by considering the random variable $-\frac{1}{k}\sum_{i=1}^k{\soc(\sd(\I,\pi^i))}$ instead of $\frac{1}{k}\sum_{i=1}^k{\soc(\sd(\I,\pi^i))}$.
\end{proof}

We are now ready to prove our main statement for \Cref{alg:approx_sc_of_rsd}.

\begin{theorem}\label{thm:sc_rsd_upper_bound}
    Let \(n \geq 1\) be an integer and \(\delta, \varepsilon \in (0, 1]\).
    For \(k \geq \frac{4n^3}{\epsilon^2}\) and \(\lambda \geq \frac{4}{\ln \nicefrac{4}{e}} \ln \frac{2}{\delta}\), the output of \Cref{alg:approx_sc_of_rsd} when applied to the assignment instance \(\I\) with \(n\) agents/items in the metric cost setting is an \(\varepsilon\)-approximation to the expected social cost of RSD with probability at least \(1 - \delta\).
\end{theorem}
\begin{proof}
    We will show that the probability that $\median(\boldsymbol{\xi})$ returned by \Cref{alg:approx_sc_of_rsd} is not an upper $\varepsilon$-approximation (respectively, a lower $\varepsilon$-approximation) of \rsd\ is at most $\delta/2$.
    \Cref{thm:sc_rsd_upper_bound} then follows by applying a simple union bound.

    We first bound the probability that $\median(\boldsymbol{\xi})$ is not an upper $\varepsilon$-approximation of \rsd.
    Notice that the values \(\xi_1, \dots, \xi_{\lambda}\) computed in \cref{alg2:line2} of \Cref{alg:approx_sc_of_rsd} are independent random variables with expectation \(\rsd(\I)\).
    For $j=1, 2, \dots, \lambda$, we define \(Z_j\) as the Bernoulli random variable that indicates whether \(\xi_j\) as computed in \cref{alg2:line2} of \Cref{alg:approx_sc_of_rsd} is not an upper \(\varepsilon\)-approximation of \(\rsd(\I)\), i.e.\
    \[
        Z_j = 
        \begin{cases}
            1 & \text{if } \xi_j \geq (1 + \varepsilon) \cdot \rsd(\I) \\
            0 & \text{otherwise}
        \end{cases}\, .
    \]
    By \Cref{lem:sc_rsd_single_instance}, we have that \(\Pr\left[ Z_j = 1 \right] \leq \tfrac{1}{4}\), and hence the random variable \(Z = \sum_{j = 1}^{\lambda} Z_j\) indicating the number of \(\xi_j\) values that are not upper \(\varepsilon\)-approximations of \(\rsd(\I)\) has expectation at most \(\tfrac{\lambda}{4}\).
    Then, the \(\median(\boldsymbol{\xi})\) exceeds \((1 + \varepsilon)\cdot \rsd(\I)\) when the random variable \(Z\) has value at least \(\tfrac{\lambda}{2}\).
    We obtain that
    \[
        \Pr\left[ \median(\boldsymbol{\xi}) \geq (1 + \varepsilon) \cdot \rsd(\I)\right]
        = \Pr\left[ Z \geq 2 \mathbb{E}[Z] \right]
        \leq \left( \frac{e}{4} \right)^{\frac{\lambda}{4}}
        \leq \frac{\delta}{2}\, ,
    \]
    as desired.
    The first inequality follows by applying the Chernoff bound (\Cref{lem:chernoff}) for the random variable \(Z\) (recall that the random variables \(Z_1, \dots, Z_{\lambda}\) are independent) with \(\eta = 1\) and the last inequality follows due to the choice of \(\lambda\).

    To bound the probability that $\median(\boldsymbol{\xi})$ is not a lower $\varepsilon$-approximation of \rsd\ is almost identical; the only change required is in the definition of the random variable $Z_j$ which should use $\xi_j \leq (1-\varepsilon) \cdot \rsd(\I)$ instead.
\end{proof}

%% file: discussion.tex
We have presented a formal statement showing that earlier \#P-hardness results on computing the RSD lottery matrix in the abstract setting imply the \#P-hardness of computing both the expected social welfare and the expected social cost of assignment instances in the value and metric cost settings, respectively. Furthermore, we have presented bounds on the number of samples sufficient and necessary to approximate these expectations with simple algorithms. Even though our analysis of \Cref{alg:approx_sw_of_rsd} for the expected social welfare is asymptotically tight, for \Cref{alg:approx_sc_of_rsd} and the expected social cost there seems to be some room for improvement. We believe that such improvements can benefit from better bounds on the expectation and variance of the social cost in terms of $n$ and $\opt(\I)$, compared to the linear bound of~\cite{CFFHT16} (see also \Cref{cl:exp-soc-cost-bound}) and our polynomial bound in \Cref{lem:exp-sc-square}, respectively. Regarding extensions of the techniques, it would be interesting to consider scenarios with more items than agents and the round-robin algorithm. We believe that our analysis for the value setting carries over to this scenario but more detailed arguments are needed for the metric cost setting.  

%% file: appendix.tex
\section{A Weaker Statement for \Cref{alg:approx_sw_of_rsd} in the Value Setting}\label{sec:app:A}

We now prove an alternative statement (\Cref{thm:weaker-bound}) for \Cref{alg:approx_sw_of_rsd} in the value setting, using Hoeffding instead of Bernstein inequality. The proof is simpler but leads to a weaker bound.

\begin{lemma}[Hoeffding inequality, e.g. see~\cite{DP09}, Chapter 1]\label{lem:hoeffding}
    Let $X_1$, $X_2$, ..., $X_k$ be independent random variables so that $X_i\in [a_i,b_i]$ and let $X=\sum_{i=1}^k{X_i}$ be their sum with expectation $\E[X]=\mu$. Then, for $t>0$, it holds
    \[\Pr\left[\left|\sum_{i=1}^k{X_i}-\mu\right|\geq t\right]\leq 2\exp\left(-\frac{2t^2}{\sum_{i=1}^k{(b_i-a_i)^2}}\right)\]
\end{lemma}

\begin{theorem}\label{thm:weaker-bound}
    Let \(n\geq 1\) be an integer and \(\delta,\epsilon\in (0,1]\). For \(k \geq \tfrac{n^2}{2 \epsilon^2} \ln \tfrac{2}{\delta}\), the output of \Cref{alg:approx_sw_of_rsd}, when applied on the assignment instance $\I$ with $n$ agents/items in the value setting, is an $\epsilon$-approximation to the expected social welfare of the RSD mechanism with probability at least $1-\delta$.
\end{theorem}

\begin{proof}
    For $i\in [k]$, define $Y_i\coloneqq \sw(\sd(\I,\pi^i))$ to be the random variable denoting the social welfare of the matching returned by \rsd\ when using the random ordering $\pi^i$. Recall that $Y_i$ takes values in $[0,\opt(\I)]$ and has expectation $\rsd(I)$. Then, the probability that \Cref{alg:approx_sw_of_rsd} does not compute an $\varepsilon$-approximation of \rsd\ is 
    \begin{align*}
    \Pr\left[\left|\frac{1}{k}\sum_{i=1}^k{Y_i}-\rsd(\I)\right|\geq \varepsilon\cdot \rsd(\I)\right] &= 
    \Pr\left[\left|\sum_{i=1}^k{Y_i}-k\cdot \rsd(\I)\right|\geq k\cdot \varepsilon\cdot \rsd(\I)\right]\\
    &\leq 2\exp\left(-\frac{2k\varepsilon^2\cdot \rsd(\I)^2}{\opt(\I)^2}\right)\\
    &\leq 2\exp\left(-\frac{2k\varepsilon^2}{n^2}\right) \leq \delta,
    \end{align*}
    completing the proof. The first inequality follows by applying Hoeffding bound (Lemma~\ref{lem:hoeffding}) to the random variable $\sum_{i=1}^k{Y_i}$ with $a_i=0$ and $b_i=\opt(\I)$ for $i\in [k]$. The second inequality follows since $\opt(\I)\leq n\cdot \rsd(\I)$ by \Cref{lem:bound_approx_sw_of_rsd_with_max_sw} and the last one by the condition on $k$.
\end{proof}

Inspecting the above proof and comparing with the last steps of the proof of \Cref{thm:sw_rsd_upper_bound}, we can see that, with the information that we have available here, all we can do in the second last inequality is to use the most pessimistic bound of $n^{-2}$ for the ratio $\rsd(\I)^2/\opt(\I)^2$ which appears in the probability bound. Instead, the use of the variance in the proof of \Cref{thm:sw_rsd_upper_bound} results in a better dependency of the probability bound on $\rsd(\I)/\opt(\I)$, and this allows us to save an $n$-factor on the bound for $k$.

\section{Upper Bounds for \Cref{alg:approx_sw_of_rsd} in the Metric Cost Setting}\label{sec:app:B}

We will now exploit \Cref{lem:exp-sc-square} to provide two guarantees for the outcome of \Cref{alg:approx_sw_of_rsd} in the metric cost setting. The first one (\Cref{thm:sc_rsd_upper_bound_Bernstein}) is obtained using the Bernstein inequality and, besides the use of the stronger bound on the variance, has a very similar structure to the proof of \Cref{thm:sw_rsd_upper_bound}. It is exponential in $n$, though, as expected by our lower bound in \Cref{thm:rsd_lower_bound_sc}.

\begin{theorem}\label{thm:sc_rsd_upper_bound_Bernstein}
    Let \(n\geq 1\) be an integer and \(\delta,\epsilon\in (0,1]\). For \(k\geq 4\max\left\{\frac{n^3}{\epsilon^2},\frac{2^n}{3\epsilon}\right\}\cdot \ln\frac{2}{\delta}\), the output of \Cref{alg:approx_sw_of_rsd}, when applied on the assignment instance $\I$ with $n$ agents/items in the metric cost setting, is an $\epsilon$-approximation to the expected social cost of the RSD mechanism with probability at least $1-\delta$.
\end{theorem}

\begin{proof}
We will use a result of~\cite{KP93} (see also~\cite{CFFHT16}), who proved that serial dictatorship always computes a matching of cost at most \(2^{n}\cdot \opt(\I)\).\footnote{The bound proved in the two cited papers is $(2^n-1)\cdot \opt(\I)$. The slightly inferior bound we use here aims to keep our calculations as simple as possible.} This means that, for $i\in [k]$ and a uniformly random ordering \(\pi^i\) of the agents in $[n]$ used by \Cref{alg:approx_sw_of_rsd}, the random variable \(\soc(\sd(\I,\pi^i))\) takes values in \([0,2^n\cdot \opt(\I)]\) and has expectation $\rsd(\I)$. By \Cref{lem:exp-sc-square}, we have that the variance of the random variable $\soc(\sd(\I,\pi^i))$ is
    \begin{align}\nonumber
        \sigma^2(\soc(\sd(\I,\pi^i))) &=\E\left[(\soc(\sd(\I,\pi^i))-\rsd(\I))^2\right]\\\label{eq:bound-variance-of-SD_sc}
        &\leq \E\left[\soc(\sd(\I,\pi^i))^2\right]\leq n^3\cdot \opt(\I)^2.
    \end{align}
    Now, for \(i\in [k]\), define the random variable $Z_i$ as \(Z_i\coloneqq\soc(\sd(\I,\pi^i)) - \rsd(\I)\). We have that the probability that the output of \Cref{alg:approx_sw_of_rsd} is not an $\epsilon$-approximation of $\rsd(\I)$ is
    \begin{align}\label{eq:bad-approximation_sc}
        \Pr\left[\left|\frac{1}{k}\sum_{i=1}^k{\soc(\sd(\I,\pi^i))} - \rsd(\I) \right| \geq \epsilon \cdot \rsd(\I)\right]
        &=\Pr \left[\left|\sum_{i=1}^k{Z_i}\right| \geq \epsilon k \cdot \rsd(\I)\right].
    \end{align}
    To complete the proof, we will bound the RHS of \Cref{eq:bad-approximation_sc}. By the definition of the random variable $Z_i$, its variance is equal to the variance of the random variable $\soc(\sd(\I,\pi^i))$, i.e.\ by \Cref{eq:bound-variance-of-SD_sc}, it holds that
    \[
        \sigma^2(Z_i) \leq n^3\cdot \opt(\I)^2.
    \]
    Furthermore, \(Z_i\) takes values in \([-2^n\cdot \opt(\I), 2^n\cdot \opt(\I)]\) and has expectation $0$.

    We now apply \hyperlink{lem:bernstein_inequality}{Bernstein inequality} (\Cref{lem:bernstein_inequality}) for the random variable \(\sum_{i=1}^k{Z_i}\) using \(t=\epsilon k\cdot \rsd(\I)\) and \(\alpha=2^n\cdot \opt(\I)\). By the discussion above, we have \(\sum_{i=1}^k{\sigma^2(Z_i)}\leq k\cdot n^3\cdot \opt(\I)^2\). Thus,
    \begin{align}\nonumber
        \Pr \left[\left|\sum_{i=1}^k{Z_i}\right| \geq \epsilon k \cdot \rsd(\I)\right]
        &\leq 2 \exp \left(-\frac{3\epsilon^2 k^2 \cdot \rsd(\I)^2}{6k \cdot n^3 \cdot \opt(\I)^2 + 2\epsilon k \cdot 2^n \cdot \opt(\I) \cdot \rsd(\I)}\right)\\\label{eq:applying-Bernstein-ineq_sc}
        &\leq 2 \exp \left(-\frac{3\epsilon^2 \cdot k}{6n^3+\epsilon\cdot 2^{n+1}}\right)
        \leq 2 \exp \left(-\frac{3\epsilon^2 \cdot k}{4\max\{3n^3,\epsilon\cdot 2^n\}}\right) 
    \end{align}
    The second inequality follows since $\rsd(\I)\geq \opt(\I)$. By \Cref{eq:bad-approximation_sc} and \Cref{eq:applying-Bernstein-ineq_sc}, we conclude that for $k\geq 4\max\left\{\frac{n^3}{\epsilon^2},\frac{2^n}{3\epsilon}\right\}\cdot \ln\frac{2}{\delta}$ we have
    \[
        \Pr \left[\left|\frac{1}{k}\sum_{i=1}^k{\sw(\sd(\I,\pi^i))}-\rsd(\I)\right| \geq \epsilon \cdot \rsd(\I)\right] \leq \delta,
    \]
    as desired.
\end{proof}

Our alternative bound on $k$ is only polynomial in \(n\) but exponential in \(\ln \frac{1}{\delta}\). The proof uses Chebyshev's inequality and is actually much simpler to prove.

\begin{lemma}[Chebyshev inequality, e.g.\ see~\cite{MR95}, page 47]\label{lem:chebyshev}
    Let \(X\) be a random variable with expectation \(\mu\) and variance \(\sigma^2\).
    Then, for \(t > 0\), it holds
    \[
        \Pr\left[|X - \mu| \geq t\right] \leq \frac{\sigma^2}{ t^2}\, .
    \]
\end{lemma}

\begin{theorem}\label{thm:sc_rsd_upper_bound_Chebyshev}
    Let $n\geq 1$ be an integer and $\delta,\epsilon\in (0,1]$. For $k\geq \frac{n^3}{\epsilon^2\delta}$, the output of \Cref{alg:approx_sw_of_rsd} when applied on the assignment instance $\I$ with $n$ agents/items in the metric cost setting is an $\epsilon$-approximation to the expected social cost of $\rsd$ with probability at least $1 - \delta$. 
\end{theorem}

\begin{proof}
    By \Cref{lem:exp-sc-square}, for \(i\in [k]\), the random variable \(\soc(\sd(\I,\pi^i))\) has variance at most \(n^3\cdot \opt(\I)^2\). Thus, the random variable \(\frac{1}{k}\sum_{i=1}^k{\soc(\sd(\I,\pi^i))}\) has expectation \(\rsd(\I)\) and variance \(\frac{n^3}{k}\cdot \opt(\I)^2\) due to the independence of the random orderings $\pi^i$ for $i\in [k]$. By applying Chebyshev's inequality (\Cref{lem:chebyshev}) to the random variable \(\frac{1}{k}\sum_{i=1}^k{\soc(\sd(\I,\pi^i))}\) with \(t=\epsilon\cdot \rsd(\I)\) and using the condition on \(k\) as well as the fact that \(\opt(\I)\leq \rsd(\I)\), we get
    \begin{align*}
        \Pr\left[\left|\frac{1}{k}\sum_{i=1}^k{\soc(\sd(\I,\pi^i))}-\rsd(\I)\right|\geq \epsilon\cdot \rsd(\I)\right]
        &\leq \frac{n^3\cdot \opt(\I)^2}{k\cdot \epsilon^2\cdot \rsd(\I)^2}\leq \delta \, .
    \end{align*}
    Hence, the probability that the output of \Cref{alg:approx_sw_of_rsd} is an $\epsilon$-approximation to $\rsd(\I)$ is at least $1-\delta$, as desired.
\end{proof}